# Strain Engineering in Monolayer $WS_2$ and $WS_2$ Nanocomposites

Fang Wang[1,2], Suhao Li[1], Mark A. Bissett[1], Ian A. Kinloch[1], Zheling Li[1]* and Robert J. Young[1]*

[1]Department of Materials and National Graphene Institute, University of Manchester, Manchester M13 9PL, UK
[2]Present address: MOE Key Laboratory of Macromolecular Synthesis and Functionalization, Department of Polymer Science and Engineering, Key Laboratory of Adsorption and Separation Materials & Technologies of Zhejiang Province, Zhejiang University, 38 Zheda Road, Hangzhou, 310027, PR China

*E-mail: zheling.li@manchester.ac.uk; robert.young@manchester.ac.uk

**Abstract**

There has been a massive growth in the study of transition metal dichalcogenides (TMDs) over the past decade, based upon their interesting and unusual electronic, optical and mechanical properties, such as tuneable and strain-dependent bandgaps. Tungsten disulfide ($WS_2$), as a typical example of TMDs, has considerable potential in applications such as strain engineered devices and the next generation multifunctional polymer nanocomposites. However, controlling the strain, or more practically, monitoring the strain in $WS_2$ and the associated micromechanics have not been so well studied. Both photoluminescence spectroscopy (PL) and Raman spectroscopy have been proved to be effective but PL cannot be employed to characterise multilayer TMDs while it is difficult for Raman spectroscopy to reveal the band structure. In this present study, photoluminescence and Raman spectroscopy have been combined to monitor the strain distribution and stress transfer of monolayer $WS_2$ on a flexible polymer substrate and in polymer nanocomposites. It is demonstrated that $WS_2$ still follows continuum mechanics on the microscale and that strain generates a non-uniform bandgap distribution even in a single $WS_2$ flake through a simple strain engineering. It is shown that these flakes could be useful in optoelectonic applications as they become micron-sized PL emitters with a band gap that can be tuned by the application of external strain to the substrate. The analysis of strain distributions using Raman spectroscopy is further extended to thin-film few-layer $WS_2$ polymer nanocomposites where it is demonstrated that the stress can be transferred effectively to $WS_2$ flakes. The relationship between the mechanical behaviour of single monolayer $WS_2$ flakes and that of few-layer flakes in bulk composites is investigated.

Keywords: micromechanics, electromechanics, transition metal dichalcogenide, photoluminescence, Raman spectroscopy

## 1. Introduction

The past two decades have witnessed an increasing interest in the application of 2D materials in various applications ranging from device fabrication to polymer composites. As one of the first 2D materials to be isolated graphene, in particular, has been studied extensively [1]. Graphene is, however, an electrical conductor that limits its wider applications where a bandgap is needed. Hence, TMDs have started to attract attention since their tunable bandgap offers more controllability [2, 3]. As a typical example of TMDs, tungsten disulfide ($WS_2$) has received significant interest for applications in transistors [4], photo-detectors [5], photovoltaic devices [6] and composites [7]. Of particular interest is that $WS_2$ exhibits a transition from a direct- to indirect-bandgap semiconductor both as the number of layers increases [8] and when the $WS_2$ flakes are subjected to strain [9]. Beyond the electronic applications, $WS_2$ flakes have also been found to have a reasonable interfacial interaction with polymers as reflected by their good reinforcement of polymers, determined by their lateral size and its distribution [10], even at a low loadings [10, 11]. This sets the foundation for making use of $WS_2$ for the next generation multifunctional nanocomposites for a number of different applications, such as transistors, sensors, photo-detectors, photovoltaics and absorbers etc. [4-6, 12].

In spite of the extensive work on the applications of $WS_2$, however, relatively little is known about one fundamental core aspect of their behaviour – the deformation micromechanics of $WS_2$ flakes for strain engineering in nanocomposites. For example, although the Young's modulus of $WS_2$ has been measured by Liu *et al.*[13] to be ~270 GPa, its bilayer exhibits a lower stiffness indicating 25% reduction of stress transfer from one layer to the other due to weak interlayer interaction, similar to that observed in few-layer graphene [14]. The state of strain at the interface between TMDs and polymers, used either as coating in devices or as matrices in thin film composites [15], is also crucial. This is because it is a fundamental aspect of deformation mechanics to ensure that no mechanical failure occurs as well as the core aspect of bandgap control. This highlights the importance of strain engineering, particularly in the monitoring of the local strain in $WS_2$ flakes and the state of stress at the interface.

Similar to earlier studies upon graphene and graphene oxide [16, 17], Raman spectroscopy has been used to follow the deformation of $WS_2$ and its Raman bands have been found to undergo a red-shift under tensile strain. In particular, the band widths and intensities are also found to correlate with the level of applied strain [18]. It is difficult, however, for Raman spectroscopy to be used to study the bandgap structure of $WS_2$. Instead, photoluminescence (PL) has recently been employed for this purpose in TMDs and it has been found the PL peaks can be used to reveal the structural non-uniformity caused by strain and doping of $MoS_2$, for which the energies of the PL peaks undergo a red shift with tensile strain [9, 19, 20]. This phenomenon was then applied to flexible $WS_2$ devices to monitor structural uniformity and strain relaxation during cyclic loading [21]. This behaviour is advantageous as PL simultaneously senses the strain as well as monitoring the bandgap in $WS_2$, suggesting a method of varying the bandgap in a single $WS_2$ flake by simple strain engineering. A recent study [22] has shown that it is possible to tailor the PL of TMD monolayers by forming hydrocarbon-filled bubbles of monolayer TMDs on a substrate. These bubbles localise the PL by producing micrometre-sized strain gradients which form so-called "artificial atoms" that are well separated on the substrate. A downside of the PL approach, however, is also apparent since, unlike in Raman spectroscopy, the intensity of PL peaks decreases greatly as the number of layers increases so that most of the PL studies to date have so far concentrated mainly on TMD monolayers [23].

In the present study, we have combined Raman and photoluminescence spectroscopy to monitor the strain distribution in mechanically-cleaved monolayer $WS_2$ flakes deformed on flexible substrates. This methodology is further extended to $WS_2$/poly(vinyl alcohol) nanocomposites where the deformation micromechanics of multilayer $WS_2$ flakes within the polymer is revealed.

## 2. Experimental

### 2.1 Materials and processing

*2.1.1 Substrate preparation.* A poly(methyl methacrylate) (PMMA) beam was coated by spin coating a 600 nm layer of SU-8 (MicroChem SU-8 2000 Permanent Epoxy Negative Photoresist) as substrate in order to enhance the contrast with the substrate under the optical microscope. The coating was undertaken using a spin coater (WS-650Mz-23NPPB spin coater Laurell Technologies Corporation). The static coating procedure involved dispensing 0.5 ml of the SU-8 resin on the PMMA substrate, followed by 10 s spinning at 500 rpm with 100 rpm/s acceleration and 30 s at 2000 rpm with 300 rpm/s acceleration [24]. The specimens were then subjected to a 1 min soft bake at 95 °C and 2 min post bake at 95 °C [24].

*2.1.2 Exfoliation and transfer.* Monolayer $WS_2$ flakes were obtained by the exfoliation through the micromechanical cleavage [25] of bulk $WS_2$ crystals with an average grain size of 200 μm supplied by HQ Graphene, Groningen, the Netherlands. The bulk $WS_2$ crystals were peeled repeatedly with an adhesive tape until very thin flakes were obtained.

The flakes obtained were then transferred to the PMMA substrate with a lay of SU-8 on top by pressing the back of the adhesive film. After the transfer, monolayer flakes were located and identified using optical microscopy and a combination of Raman and photoluminescence (PL) spectroscopy.

*2.1.3 Nanocomposite preparation.* Liquid-phase exfoliation was employed to produce WS$_2$ dispersions, termed LPE WS$_2$. A detailed procedure of the exfoliation was reported earlier by Bissett *et al.* [26]. Briefly, dispersions of LPE WS$_2$ were produced by ultrasonication for 12 hours of commercially-available WS$_2$ powder (Sigma-Aldrich) in a mixture of isopropanol and water (1:1 v/v) at a concentration of 10 mg/ml [26]. This was followed by centrifugation at three different speeds (1500, 3000 and 6000 rpm) so that a range of lateral dimensions could be obtained, namely LPE WS$_2$-L, LPE WS$_2$-M and LPE WS$_2$-S, respectively. The WS$_2$/PVA thin film nanocomposites were fabricated *via* a solution-mixing method. The poly(vinyl alcohol) (PVA, Sigma-Aldrich) was firstly dissolved in deionized water at ~90 °C to form aqueous solutions (~50 mg/ml). The polymer solutions were mixed with the dispersions of WS$_2$ in the appropriate ratio with WS$_2$ to give loadings of 0.5, 0.8, 1.2, 2.0 and 5.0 wt%, respectively, followed by a short sonication (~5 min). Finally, the solutions were dried at room temperature for a few days (> 48 h) to obtain the composites. The LPE WS$_2$ loading was converted to volume fraction $V_f$ (vol%) from the weight fraction $W_f$ (wt%) using the following equation:

$$V_f = \frac{W_f \rho_p}{W_f \rho_p + (1-W_f)\rho_W} \quad (1)$$

where $\rho_p$ and $\rho_W$ are density of the polymer and WS$_2$ flakes, which were taken as 1.3 g/cm$^3$ [17] and 7.5 g/cm$^3$ [27], respectively. As a result, the 0.5, 0.8, 1.2, 2.0 and 5.0 wt% weight fraction loadings are converted to volume fractions of 0.09, 0.14, 0.21, 0.35 and 0.90 vol%, respectively.

## 2.2 Characterisation

An atomic force microscope (AFM, Bruker dimension 3100) was used in the tapping mode to provide morphological information about the WS$_2$ flakes deposited onto a Si/SiO$_2$ substrate. For transmission electron microscopy (TEM) analysis, the WS$_2$ flakes were tape exfoliated onto a Si/SiO$_2$ substrate. They were then transferred to a lacey-carbon Cu TEM grid via a polymer-free technique. The TEM phase contrast images were obtained using A JEOL 2100 field emission TEM with an accelerating voltage of 200 kV in a bright field detector. The acquisition time was between 0.5 s and 2.5 s.

The concentrations of WS$_2$ dispersions were determined using a UV-Visible spectrophotometer (Thermal Scientific Evolution 201) with a Beer's Law Coefficient of 2756 ml/mg/m (for peak at $\lambda$=629 nm) [28]. Raman and PL spectroscopy were conducted using a Horiba LabRam system using $\lambda$=488 nm laser excitation at room temperature. The laser power for both spectroscopy measurements was set at a low level to avoid laser heating of specimens. The orientation of the LPE WS$_2$ filler was examined using polarised Raman spectroscopy [29]. The *in-situ* deformation testing of the LPE WS$_2$/PVA nanocomposites was undertaken using a Renishaw system 1000 Raman spectrometer with a $\lambda$=514.5 nm laser excitation with low power (<0.2 mW). The tensile properties of the neat PVA and LPE WS$_2$/PVA composites with different WS$_2$ loadings were determined based on the ASTM D3039 method using an Instron-1122 universal testing machine. Prior to mechanical testing, the specimens were left for 24h in an air-conditioned laboratory where the temperature was set as 23.0 ± 0.1 °C with a relative humidity of ~50 ± 5%. The specimens were deformed using a crosshead speed of 1 mm/min and four or five specimens for the neat PVA sample and each of the LPE WS$_2$/PVA composites sample were tested.

## 2.3 In-situ deformation with PL and Raman spectroscopy

For deformation of the monolayer WS$_2$ flakes, the PMMA substrate was mounted on a 4-point bending rig and placed under the microscope stage in Raman spectrometer. A resistance strain gauge was fixed to the PMMA beam close to the WS$_2$ flakes to monitor the strain. The specimen was then deformed stepwise with the Raman or PL spectra collected for each strain step. Mapping was undertaken using a grid size of 0.5 μm × 0.5 μm or 1 μm × 1 μm depending upon sizes of the flakes and the degree of precision needed. The laser was polarised parallel to the direction of tensile strain. For the deformation test of thin film nanocomposites, the films were mounted and stretched in a tensile rig designed in house and Raman spectra were obtained for the LPE WS$_2$/PVA nanocomposites. The details of the experimental method have been described in detail in our previous report [7].

## 3. Results and Discussion

### 3.1 Photoluminescence spectroscopy of monolayer WS$_2$

Figure 1a shows an optical image of a typical WS$_2$ flake prepared *via* micromechanical cleavage consisting of a thick multilayer and a monolayer adjacent to it as confirmed by PL. Figure 1b shows the PL spectrum of the monolayer consisting of an extremely strong peak with two features, namely: a dominant peak centered at ~2.00 eV corresponding to a neutral *A* exciton and a weaker peak with slightly lower energy known as the Trion exciton ($A^-$ exciton) caused by undesired doping during exfoliation [30, 31]. In contrast, Figure 1c also shows that the intensity of the PL peaks reduces significantly as the number of layers of WS$_2$ increases, and the spectrum from the bulk crystal shows nearly no peak at all [23, 32]. This is a result of the WS$_2$ undergoing the transition from indirect band gap (Γ-K) to direct band gap (K-K) semiconductor as the number of layer decreases towards monolayer [23, 32]. This can therefore be used to identify monolayer $WS_2$ since only the monolayer region shows a strong intensity of the *A* exciton peak whereas there is no detectable signal from thicker counterparts (inset in Figure 1a).

The high sensitivity of PL spectroscopy to detect features at the submicron level also reveals some non-uniformity around the centre of flake where wrinkles and contaminant are present [20]. This can also be seen by the AFM (Figure 1d) The AFM height profile also shows some contamination on the flake, perhaps due to bubbles trapped by the flake, commonly seen for micromechanically-exfoliated flakes of 2D materials [22, 33]. It should be noted that the height profile along the line in Figure 1d is around 10 nm, significantly greater than the thickness of $WS_2$ monolayer. This may arise from high degree of roughness and viscoelastic nature of polymer substrate. The lattice structure of the $WS_2$ flakes can also been clearly seen by using atomic resolution TEM as shown in Figure 1e [7].

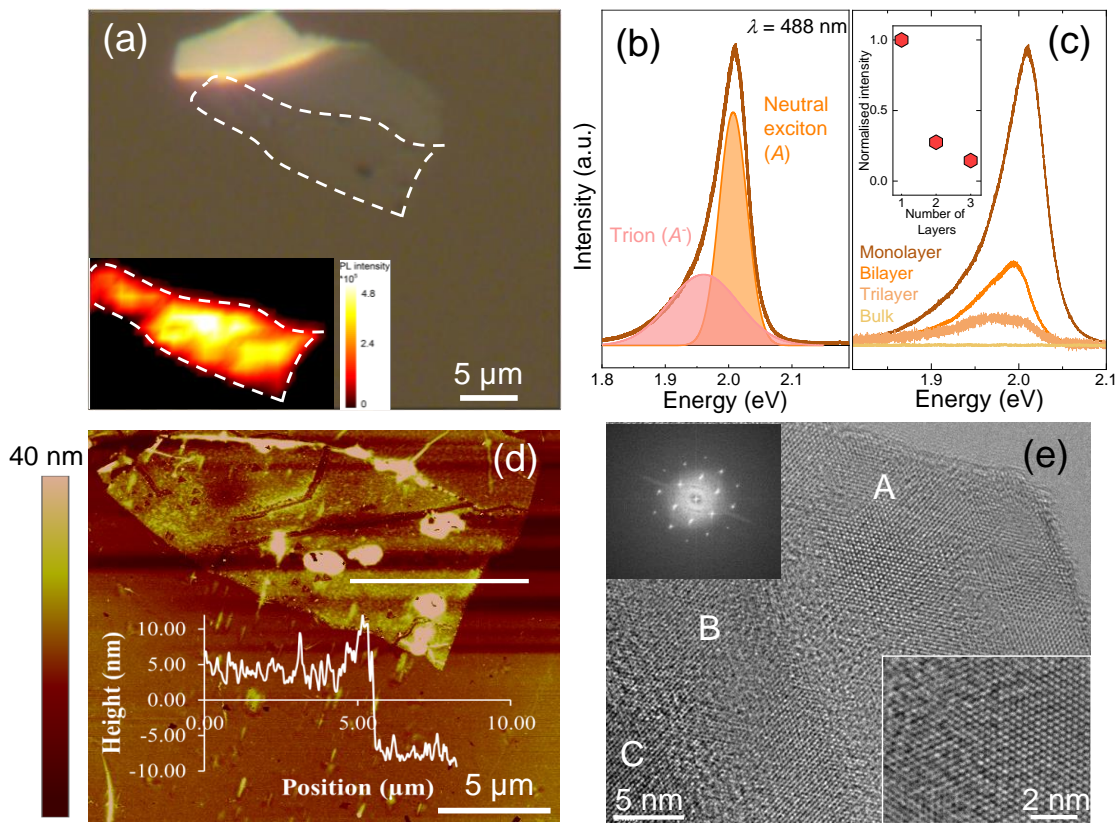

Figure 1. (a) Optical image of a monolayer $WS_2$ flake (outlined) along with a multilayer flake. The inset in (a) is the PL intensity mapping of the *A* exciton peak which only highlights the monolayer flake. (b) PL spectrum of the monolayer $WS_2$ flake in (a). (c) The PL spectra of $WS_2$ flakes with different numbers of layers. The inset shows their normalised intensity. (d) AFM micrograph of the monolayer flake in (a). The height profile along the white line in (d) is shown as inset. (e) TEM image of a $WS_2$ flake with regions of different thickness. Areas A, B, and C show the atomic structure most likely of single, double, and triple layers, respectively. The lower inset clearly indicates an atomically clean $WS_2$ lattice, and the FFT image in the top inset indicates the crystalline nature of the flake. (Image courtesy of Eoghan O'Connell and Ursel Bangert, University of Limerick, Ireland)

Figure 2a is the schematic diagram the PMMA substrate with SU-8 top coating in the 4-point bending rig placed on the stage of the Raman spectrometer [16, 34]. It should be noted that the size of $WS_2$ flake (~10 μm) is three orders of magnitude smaller than the length of substrate (~70 mm) so that the flake can be assumed to be strained uniaxially [16]. A clear red-shift of the PL peak can be seen as the strain level increases, with the peak fitted with two Gaussian peaks (Figure 2b) [7, 35]. Since the band gap in monolayer $WS_2$ is mostly determined by the 3p orbital of the S atoms and the 5d orbital of the W atoms [36, 37], the red-shift occurs as the strain alters the W-W and W-S bond lengths, giving rise to a reduction in the orbital hybridization and d-band width [38]. In more detail, as the strain increases, the red-shift of both the $A$ exciton peak and $A^-$ exciton peak is in the order of tens of meV as shown in Figure 2c, which agrees with other studies on TMDs [19, 23, 38, 39]. The shift rate is found to be -58.7 ± 1.4 meV/% strain for the $A$ exciton peak, and -89.9 ± 4.9 meV/% strain for the $A^-$ exciton peak, respectively. The apparent tension at 0% strain is likely to be due to the residual strain induced during specimen preparation. These shift rate values are similar to those predicted by theoretical simulation [40], and amongst the highest experimental values reported [35, 41].

The PL spectra $x$-$y$ mapped on a 0.5 × 0.5 μm grid across a monolayer $WS_2$ flake, in a region showing no contamination, (Figure 2d) were collected at both 0% and 0.35% strain, A reduction of the energy (bandgap) can be clearly seen towards the centre of the flake for 0.35% strain (Figure 2e). This implies that the simple uniaxial tension applied to the monolayer flake on the substrate generates a non-uniform bandgap distribution across an individual $WS_2$ flake. This is of particular relevance for local fine tuning of the bandgap in a TMD flake through strain engineering [20]. The PL energy varies by some 40 meV over a region of a few microns in the $WS_2$ monolayer crystals at 0.35% strain. Moreover, this variation of PL energy can be controlled by the application of external strain to the substrate as long as the $WS_2$ monolayer crystals remain intact and do not debond from the substrate.

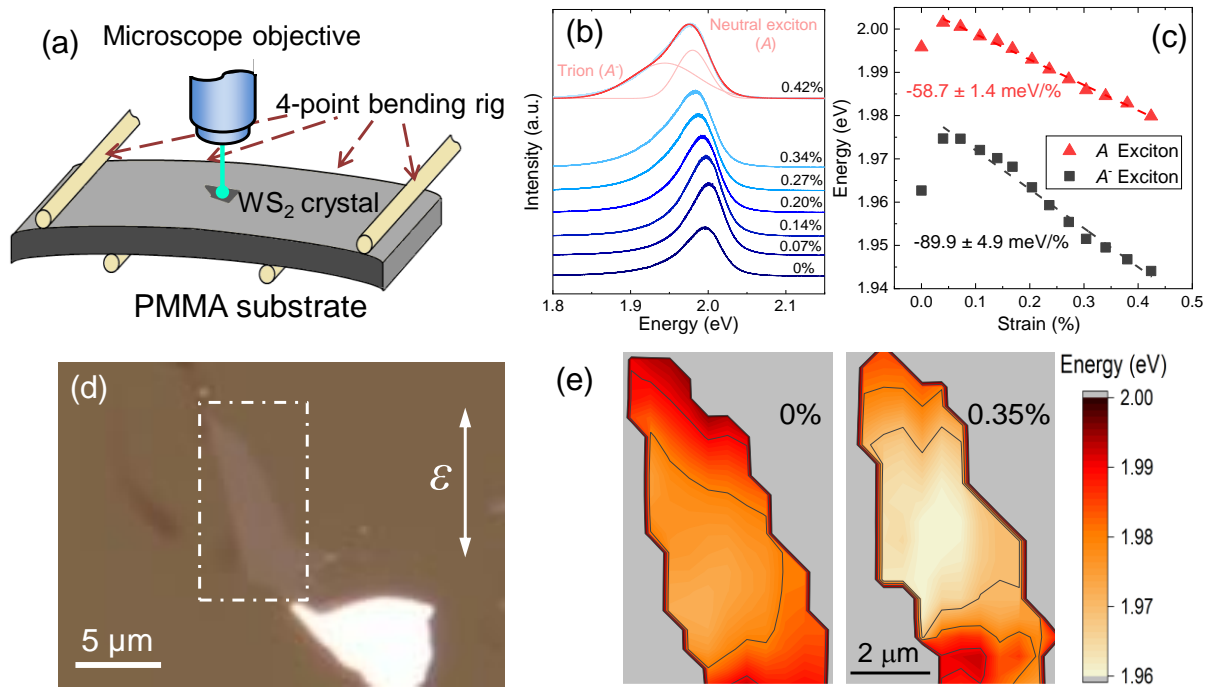

Figure 2. (a) Schematic drawing of the '*in-situ*' deformation set up in PL and Raman spectroscopy. (b) Evolution of the PL spectrum of a monolayer $WS_2$ flake as strain increases. (c) The energies of $A$ exciton and $A^-$ exciton peaks as a function of strain, with the solid lines being the linear fitting. (d) Optical image of one monolayer $WS_2$ flake. (e) Map of the $A$ exciton peak energy over the flake at (a) 0% and (b) 0.35% strain.

It is interesting to compare and contrast this present study with that of Tyurina and coworkers [22] who demonstrated that it was possible tailor the PL using hydrocarbon-filled bubbles under monolayer TMDs on a substrate. Although they were able to localise the PL variation in submicron regions ("artificial atoms") they were only able to tune the behaviour though the choice of different TMDs rather than by external strain as demonstrated in our study.

The measurements of the positions of the $A$ exciton peaks in Figure 2e were converted to the strain through the use of the correlation found in Figure 2c, leading to maps of the monolayer $WS_2$ strain distributions at different applied strain (the strain that was applied to the substrate) levels (Figure 3a & b). It can be seen that at 0% applied strain the flake has an average strain about 0% but with some variation, which could be attributed to the formation of wrinkles in the flake from the specimen preparation by pressing the flake onto the substrate [42, 43], that can be flattened by the tensile strain applied afterwards. When the applied strain is increased to 0.35%, the strain in the flake is found to increase from the edges and eventually plateau in the middle of the flake. This behaviour is shown more clearly in Figure 3c where the variations of strain along the dashed lines shown in Figures 3a & b are plotted.

The behaviour shown in Figure 3 is characteristic of that seen during the deformation of monolayer graphene on a substrate for which it has been demonstrated that continuum mechanics [44] can be employed to model strain distributions using the well-established 'shear-lag' theory [16, 45]. In this theory, the strain in the WS$_2$ flake, $\varepsilon_f$, at a position $x$ along the strain direction is given by an equation of the form [16, 46]:

$$\varepsilon_f = \varepsilon_m \left[1 - \frac{\cosh(ns\frac{x}{l})}{\cosh(ns/2)}\right], \quad n = \sqrt{\frac{2G_m}{E_f}\left(\frac{t}{T}\right)} \quad (2)$$

where $\varepsilon_m$ is the level of strain applied to the matrix, $E_f$ is the Young's modulus of the WS$_2$ flake, $G_m$ is the shear modulus of the polymer, $l$ is the length of WS$_2$ flake along with strain direction, $t$ and $T$ are the thickness of the WS$_2$ flake and the representative volume, respectively and $s$ is the aspect ratio of the WS$_2$ flake ($l/t$).

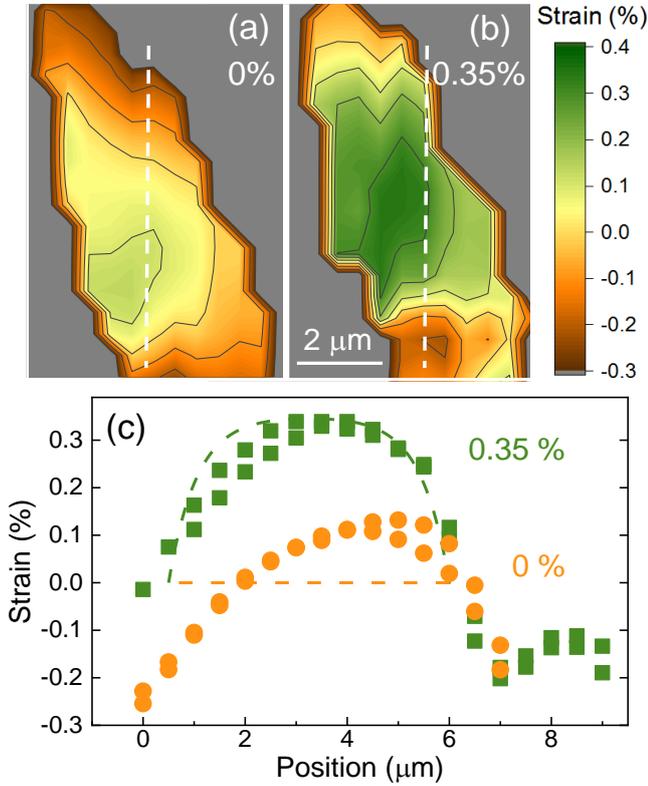

Figure 3. Strain distribution in the monolayer WS$_2$ flake measured using position of the $A$ exciton peak at (a) 0% and (b) 0.35% applied strain. (c) Extracted strain distribution in the region along the white dashed line (starting from the top) in (a) and (b). The dashed line is indicative of the strain distribution at 0% strain (orange) and 0.35% strain (green) drawn using the shear-lag analysis (Eq. 2) with $ns$=10.

A 'shear-lag' curve drawn using Eq. 2 with $ns = 10$ (Figure 3c) at an applied strain ~0.35% shows the shear-lag theory to be an effective model to predict the behaviour of WS$_2$ flake on a polymer substrate, and suggests an intact interface between monolayer and polymer. The critical length, $l_c$, which is defined as twice the distance from the flake edge (0% strain) to where 90% of the maximum strain occurs [46], has also been estimated to be in the order of ~4 µm at a strain of 0.35%, which provides an important guideline regarding the minimum lateral dimension of WS$_2$ flakes reinforcing nanocomposites. Similarly, the variation of the shear stress, $\tau$, along the WS$_2$ flake can be determined by developing Eq. 2 using a force balance [16] to give:

$$\tau = nE_f\varepsilon_m \frac{\sinh(ns\frac{x}{l})}{\cosh(ns/2)} \quad (3)$$

By using $ns = 10$, $E_f = 272$ GPa [13] and $t = 0.65$ nm [13], the shear stress at the edges of the monolayer WS$_2$ flake is found to be ~1.1 MPa. This value is similar to that found for monolayer graphene [16, 47] but is an order of magnitude lower than that of the carbon fibres in composites (~20-40 MPa) [46].

Although found previously for graphene [16], graphene oxide [45] and MoS$_2$ [20], it is shown previously [44] and in this work that monolayer WS$_2$ also follows continuum mechanics at the microscale. Hence it seems that this behaviour may be typical for all 2D layered materials. In particular, it has been demonstrated that the non-uniformity of strain determines the distribution of the bandgap, as monitored by the energy of PL peak. This suggests a way of tuning the bandgap of WS$_2$ even within a single flake by making use of the non-uniformity in strain-engineered TMDs.

A recent study [48] has reviewed the experimentally-determined PL shift rates for a number of TMDs, including WS$_2$. Values of shift rate ranging from as low as -1.3 meV/% [49] to -61.8 ± 3.8 meV/% [50] have been reported for the $A$ exciton peak of WS$_2$. The shear lag analysis described above enables this discrepancy to be explained. The stress transfer efficiency is controlled by the parameter $ns$ and the interaction of the WS$_2$ flake with the substrate [51]. Better stress transfer is given by high values of $ns$, i.e. using stiffer substrates (with higher $G_m$ values leading to a higher $n$) and long flakes with high aspect ratios $s$. For example the low value of shift rate was found using a very flexible polydimethylsiloxane (PDMS) substrate [49]. This present study used long monolayer flakes on a relatively-stiff PMMA substrate with a layer of SU-8 top coating.

### 3.2 Raman spectroscopy of WS$_2$ flakes

As shown above, the intensity of the PL peaks in WS$_2$ drops significantly as the number of layers increases. This limits its wider application in fields where few-layer and multilayer TMDs are employed, such as in nanocomposites [15]. In contrast, the intensity of the Raman bands in WS$_2$ tends to increase as the number of layers increases and so Raman spectroscopy can therefore, be employed both as a complementary technique for strain measurement and for comparison purposes.

Monolayer and few-layer WS$_2$ flakes were located on a substrate (Figure 4a) and their Raman spectra are shown in Figure 4b. The monolayer was confirmed by the characteristic strong PL emission (inset in Figure 4b) and, in the Raman spectra, the increase of number of layers results in the increasing separation between $E_{2g}^1$ and $A_{1g}$ bands (~62 cm$^{-1}$ for monolayer) [52]. As the tensile strain is increased up to ~0.55%, both the $E_{2g}^1$ and $A_{1g}$ bands of the monolayer and few-layer WS$_2$ not only undergo a detectable red-shift, but also show a splitting of the $E_{2g}^1$ band into two bands, namely the $E_{2g}^{1-}$ and $E_{2g}^{1+}$ mode, respectively (Figure S1). This arises from the removal of the degeneracy of the $E_{2g}^1$ mode as a result of strain [35, 39]. As the $E_{2g}^{1+}$ band stays nearly unshifted, we will now demonstrate how the use of the Raman $E_{2g}^{1-}$ mode in a similar way to our previous report [7] enables the strain in WS$_2$ flake to be determined. (The band will be referred to as the $E_{2g}^1$ mode in the following discussion for simplicity).

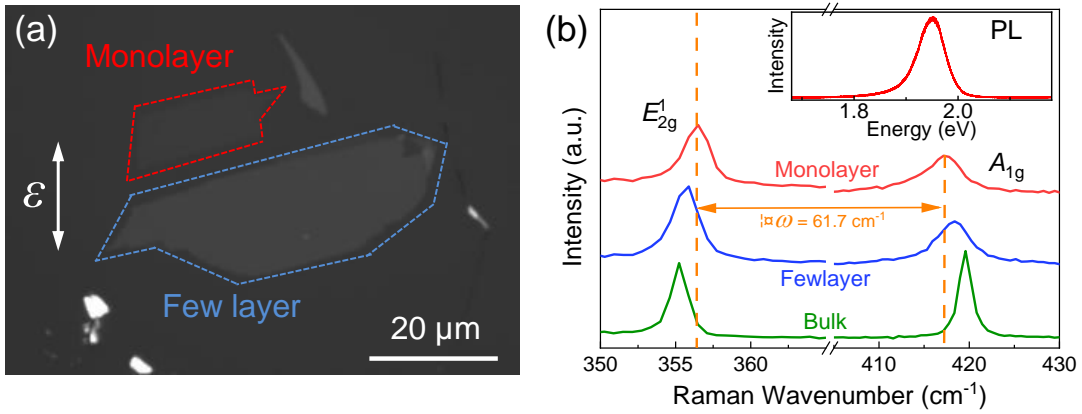

Figure 4. (a) Optical micrograph of monolayer and few layer WS$_2$ flakes. (b) Raman spectra of monolayer, few-layer and bulk WS$_2$ flakes. The inset shows the PL spectrum of a monolayer WS$_2$ flake.

In order to investigate the micromechanics of monolayer WS$_2$ flake using Raman spectroscopy, the PMMA beam was deformed using the procedure summarised in Figure S2. Briefly, the beam with monolayer WS$_2$ on top was deformed initially to a strain ~0.55%. The specimen was then released and top-coated with a thin layer of SU-8 photoresist polymer. In the second cycle the specimen was then loaded to the same strain of ~0.55%. Using the calibration between the Raman $E_{2g}^1$ band and strain established in our previous study [7], the strain distributions of the uncoated monolayer WS$_2$ at 0% and 0.55% applied strain were mapped using a 1 μm × 1 μm grid as shown in Figure 5a and b, respectively.

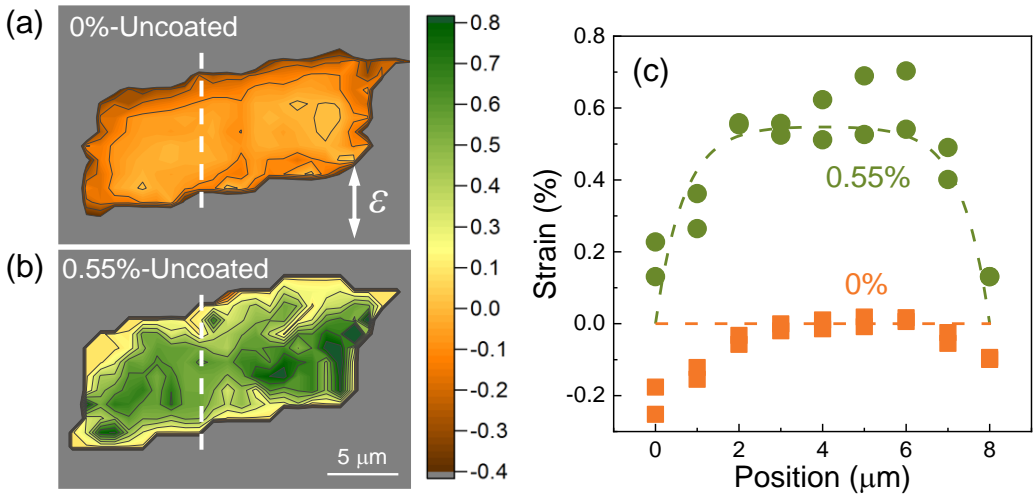

Figure 5. Strain distribution of uncoated monolayer WS$_2$ flake at (a) 0% and (b) 0.55% applied strain. (c) Extracted strain distribution in the region along the white dashed lines in (a) and (b) determined from the frequency of $E_{2g}^{1-}$ mode. The dashed lines are indicative of strain distribution at 0% and 0.55% strain drawn using the shear-lag analysis Eq. 2 with $ns=12$.

It can be seen that the strain distribution in Figure 5 is quite uniform at 0% strain, apart from the compression at the edge of flake perhaps due to sample preparation. The similarity between the strain distributions shown in Figure 5 obtained using Raman spectroscopy and those obtained for a similar WS$_2$ monolayer specimen using PL in Figure 3 is striking.

When the specimen was subjected to a strain of 0.55%, the strain in most of the WS$_2$ flake increased to about 0.55%, demonstrating effective stress transfer from the substrate to the flake. However, a slight strain concentration, higher than the applied strain ~0.55%, can be found in the bottom part perhaps due to defects or edge effects [53]. The data points extracted along the vertical dashed white lines in tensile strain direction (Figure 5c) clearly show a strain plateau in the middle of the flake at the strain of 0.55%, suggesting that monolayer WS$_2$ flake also follows the shear lag behaviour shown in Figure 3. Similarly, by using $ns$ = 12 in Eq. 2, the 'shear-lag' curve is drawn for an applied strain ~0.55% (Figure 5c). Similar to the behaviour shown in Figure 3d, the curve matches the data points very well, further confirming the validity of shear-lag theory. The shear stress at the edges of the monolayer WS$_2$ flake is found to be ~1.5 MPa (Eq.3), similar to that found by using PL (Figure 3).

The specimen shown in Figure 5 was unloaded and a top coat of SU-8 applied so that the deformation of the WS$_2$ monolayer sandwiched between two polymer films, as in a nanocomposite, could be studied as shown in Figure 6. Although the strain in the monolayer is relatively uniformly distributed at 0% strain (Figure 6a) it is, however, reduced to ~-0.35% which means that there is a uniform 0.35% compressive strain in the flake as a result of the shrinkage of the top-coating SU-8 polymer during curing [54]. Subsequently, the coated flake was subjected to a strain of 0.55% (Figure 6b) and a strain concentration at top part of the flake is observed. The 'shear-lag' curve drawn using Eq. 2 ($ns$=12) for an applied tensile strain of 0.55% is offset by -0.35% to take into account the 0.35% compressive strain, hence the maximum strain is only about 0.20% [53]. In addition the strain drops sharply in the middle of the flake although the strain distribution of other parts still approximately follows the shear lag curve (Figure 6c). The reason for this is attributed to the occurrence of two cracks in the middle of the WS$_2$ flake perpendicular to the tensile axis as highlighted with dashed white frames and the inset in Figure 6b, similar to the fracture behaviour observed for monolayer graphene [55]. This implies that the failure strain of this WS$_2$ flake is no more than 0.55%, due probably to the presence of defects. This strain to failure, in combination with its Young's modulus ~272 GPa [13], implies the strength of this monolayer WS$_2$ flake was ~1.5 GPa. The interfacial adhesion has not been improved by the SU-8 top coating. This is because for monolayer WS$_2$ sandwiched and stretched between SU-8 layers, the adhesion with one side interface is not significantly different from that with double sides. This is in similarity with graphene, but if few-layer graphene is stretched the SU-8 top coating can provide extra adhesion to compensate for the interlayer sliding due to the weak interlayer bonding [14].

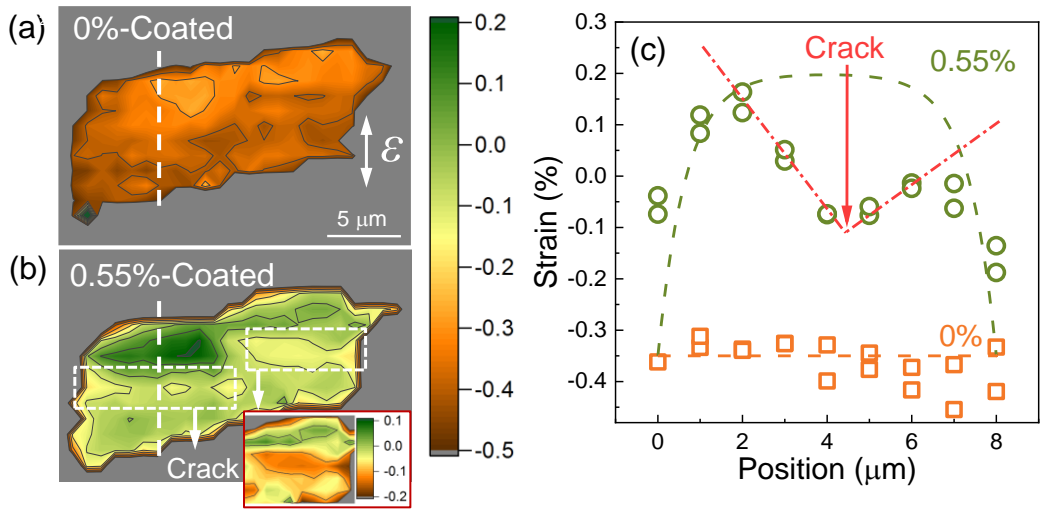

Figure 6. Strain distribution of coated monolayer WS$_2$ flake at (a) 0% and (b) 0.55% applied strain. In (b) the location of crack is also indicated. The right crack is highlighted as the inset with a finer strain scale. (c) Extracted strain distribution in the region along the white dashed lines in (a) and (b) determined from the Raman frequency of $E_{2g}^{1-}$ mode. The dashed lines are indicative of strain distribution at 0% and 0.55% strain drawn using the shear-lag analysis in Eq. 2 with $ns$=12.

## 3.3 Thin film $WS_2$/PVA nanocomposites

We extended our analysis of the deformation micromechanics of monolayer $WS_2$ further to that of bulk few-layer $WS_2$/PVA nanocomposites. The few-layer $WS_2$ flakes were liquid-phase exfoliated in large quantities and separated by centrifugation into three grades based on their lateral dimensions, namely, large (LPE $WS_2$-L), medium (LPE $WS_2$-M) and small (LPE $WS_2$-S) $WS_2$ flakes [56]. The lateral size distributions of each grade, measured from SEM images for more than 60 flakes of each grade, are presented in Figure 7a. It can be seen that LPE $WS_2$-L, LPE $WS_2$-M and LPE $WS_2$-S have average lateral sizes of 8.1 ± 4.7 µm, 2.4 ± 2.1 µm and 1.0 ± 0.5 µm, respectively.

Raman spectra of the LPE-$WS_2$ flakes, neat PVA and their nanocomposites are shown in Figure 7b, with both the $E_{2g}^1$ and $A_{1g}$ bands can be clearly identified in the nanocomposites. The position of the $E_{2g}^1$ band ($\omega(E_{2g}^1)$) was used to monitor the strain in the $WS_2$ flakes during deformation of the nanocomposites containing the few-layer $WS_2$ flakes of different lateral dimensions. It can be seen in Figure 7c that the nanocomposites reinforced with LPE $WS_2$-L flakes exhibit a significant red-shift at a rate of -0.66 ± 0.15 cm$^{-1}$/% strain (Raman spectra shown in Figure S3), while the shift rate decreases as the lateral dimension decreases because of lower level of reinforcement provided by the smaller flakes [34]. It was found that the rate of band shift decreased $WS_2$-L/PVA above ~0.3% strain due possible to failure of the interface between the PVA and $WS_2$ flakes above this strain [57]. This is further evidenced by similar behaviour of the $A_{1g}$ band (Figure S4) where the LPE $WS_2$–L flakes show significantly larger shift than the other two. The results were found to be very reproducible, in that only the LPE $WS_2$–L flakes with large lateral dimensions of 8.1 ± 4.7 µm give rise to good stress transfer and so can be used to reinforce nanocomposites effectively.

The critical length for the monolayer $WS_2$ flakes was found to be ~4 µm and this is indicated on Figure 7a. It should be noted that the lateral sizes of most of the LPE $WS_2$–L flakes are higher than 4 µm whereas they are lower than 4 µm for the LPE $WS_2$–M and LPE $WS_2$–S flakes. This is clear indication of the importance of flake size in having good stress transfer in nanocomposites [58]. It is likely that the larger flakes may also be thicker and it might be better to consider the flake aspects ratios, $s$, rather than just their lateral dimensions for a complete analysis in the future.

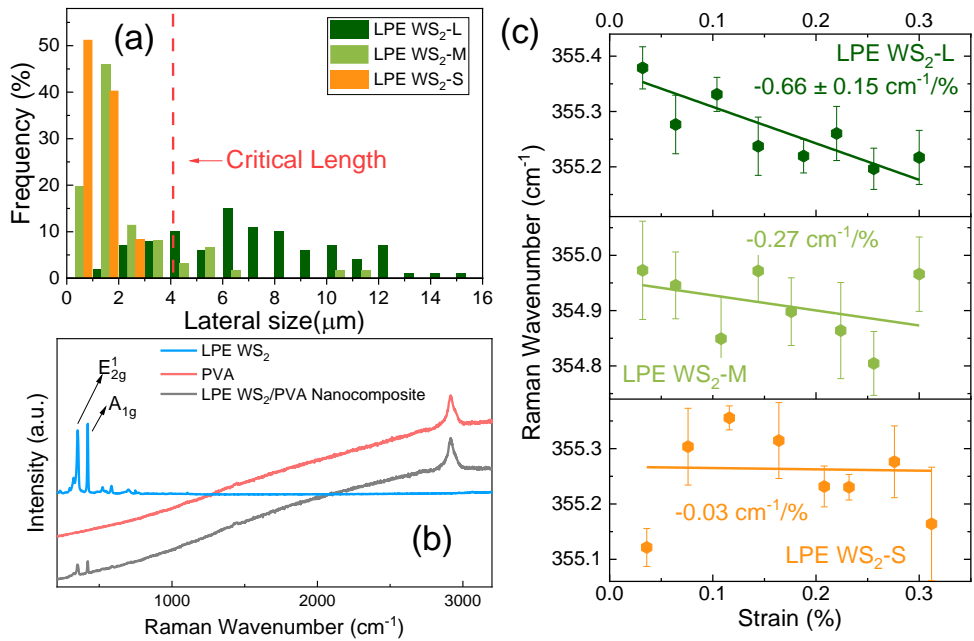

Figure 7. (a) Distribution of lateral size of the LPE $WS_2$-L, LPE $WS_2$-M and LPE $WS_2$-S flakes. (b) Raman spectra of the LPE $WS_2$ flakes, neat PVA and their nanocomposites. (c) Shift of Raman $E_{2g}^1$ band as the function of strain of nanocomposites reinforced by LPE $WS_2$-L, LPE $WS_2$-M and LPE $WS_2$-S, respectively. The straight lines are the linear fit of the corresponding data points.

The Raman band shift rate reveals the level of strain in the materials as a result of stress transfer from matrix and the stress/strain-induced Raman band shift and has been used to estimate the effective modulus of graphene [59], graphene oxide [17] and boron nitride [60] in nanocomposites. We have made an attempt to extend this methodology to WS$_2$ flakes [45]. The effective modulus of the reinforcement is given by

$$E_{\text{eff}} = \frac{d\omega(E_{2g}^1)}{d\varepsilon} \cdot \frac{E_f}{d\omega(E_{2g}^1)/d\varepsilon(\text{ref})} \quad (4)$$

where $d\omega(E_{2g}^1)/d\varepsilon$ is the shift rate of the Raman $E_{2g}^1$ band as the function of strain $\varepsilon$. The reference band shift rate value, $d\omega(E_{2g}^1)/d\varepsilon(\text{ref})$ for the deformation of isolated flakes can be taken as -2.05 cm$^{-1}$/% as found previously [7]. If the Young's modulus of a monolayer WS$_2$ flake $E_f$ is taken as 272 GPa [13], then the value of $d\omega(E_{2g}^1)/d\varepsilon$ measured -0.66 ± 0.15 cm$^{-1}$/% for LPE WS$_2$–L flakes in this work (Figure 7c) leads to an effective modulus of LPE WS$_2$-L ($E_{\text{eff}}$) of ~87 ± 20 GPa, i.e. around 30% of the value of 272 GPa determined from the direct measurement on monolayer flakes [13]. No apparent slipping in either model devices or nanocomposites has been observed in PL and Raman experiments. This is for two reasons: (1) the level of strain achieved is not sufficiently high to induce an interfacial damage and (2) the WS$_2$ flake cracks first prior to the interfacial failure, probably due to the fact that the strength of WS$_2$ is not as high as graphene, especially with the presence of defects [55].

We also investigated the reinforcement of the WS$_2$/PVA nanocomposites further by undertaking tensile testing upon the nanocomposites reinforced with different loadings of the LPE WS$_2$-L flakes. The stress-strain curves in Figure 8a show that the incorporation of WS$_2$ at low loadings (up to ~0.21 vol%) there is a linear increase in the Young's modulus of the nanocomposites ($E_c$) (Figure 8). Above this loading, the reinforcement efficiency from the WS$_2$ decreases, probably due to aggregation and percolation effects [61], although the maximum $E_c$ is found to be ~940 ± 38 MPa for the highest loading, corresponding to nearly a 60% increase in modulus with respect to the neat PVA. In contrast, the strain to failure and ultimate strength remain unchanged, regardless of the WS$_2$ loading (Figure 8a). These findings are consistent with the Raman band shifts shown in Figure 7c.

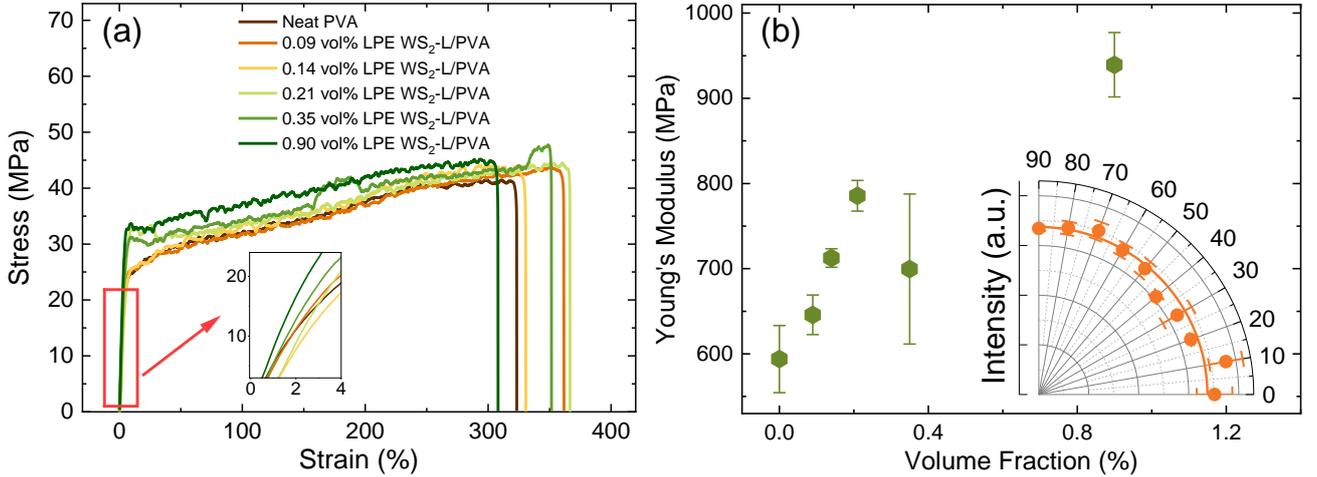

Figure 8. (a) Stress-strain curves for the neat PVA and LPE WS$_2$-L/PVA nanocomposites with different LPE WS$_2$-L flake loadings. (b) Young's modulus of the neat PVA and nanocomposites with different LPE WS$_2$-L loadings. The inset is the intensity of Raman $A_{1g}$ band as the function of rotation angle.

We have demonstrated previously that the Young's modulus of the nanocomposites, $E_c$, can be analysed using the modified rule of mixtures [46, 58]:

$$E_c = E_{\text{eff}} V_f + E_m(1-V_f) \quad (5)$$

where $E_{\text{eff}}$ is the effective Young's modulus of LPE-WS$_2$, $E_f$ and $E_m$ are the Young's modulus of the LPE-WS$_2$ and PVA, respectively. $V_f$ is the volume fraction of WS$_2$ in the nanocomposites. Figure 8b shows a linear increase of $E_c$ as the function of $V_f$, up to 0.21 vol% loading that enables the $E_{\text{eff}}$ to be determined as ~93 GPa, similar to that obtained from Raman band shifts using Eq. 4, validating the use of Raman spectroscopy to predict the value of $E_{\text{eff}}$ for the TMDs.

The effective modulus of WS$_2$ flakes in the nanocomposites is given by [58]

$$E_{\text{eff}} = \eta_l \eta_o E_f \quad (6)$$

The length factor, $\eta_l$, which has a value of between 0 and 1, reflects the dependence of reinforcement on flake length and increases with the flake aspect ratio $s$ [62]. The Krenchel orientation factor $\eta_o$ enables the effect of filler orientation upon the reinforcement efficiency to be determined and it ranges from 8/15 for randomly oriented to 1 for well-aligned flakes [29, 63]. Polarised Raman spectroscopy (inset in Figure 8b) shows that the intensity of the $A_{1g}$ band is independent of the angle of direction of laser polarisation. This corresponds to a random orientation of the $WS_2$ flakes in the nanocomposites, based on the methodology established earlier [29]. The length factor $\eta_l$ is given by [64]

$$\eta_l = 1 - \frac{\tanh(ns/2)}{ns/2} \quad (7)$$

The parameter $ns$ was found to be of the order of 10 for the monolayer flakes which would give value of $\eta_l \sim 0.8$ (Eqs. 2 and 7) as the average number of $WS_2$ layers is about 1-3 [26]. Considering the above, the values of $E_f$ can be calculated to be up to 220 GPa according to Eq. 6, in broad agreement with the experimental theoretical value ~270 GPa. It should be noted that the value of $E_f$ may also be lower for few-layer materials than the monolayer, as has been found for graphene [14], due to interlayer sliding.

## 4. Conclusions

In this work, photoluminescence and Raman spectroscopy have been combined to successfully monitor the strain distribution and stress transfer of monolayer $WS_2$ on a flexible polymer substrate and also in thin film bulk nanocomposites. It is demonstrated that monolayer $WS_2$ still follows continuum mechanics. Particularly, a non-uniform bandgap distribution has been achieved by strain engineering even within a single $WS_2$ flake due to the non-uniform strain distribution through stress transfer from the substrate. It has been demonstrated that this could have useful applications in optoelectonics in producing tuneable micron-sized PL emitters on a substrate. The micromechanics developed for monolayer $WS_2$ flake has been extended to thin film nanocomposites, where it has been found that the stress can be transferred effectively to thicker $WS_2$ flakes. Their effective Young's modulus is around 30 % of their theoretical modulus which means that their reinforcement efficiency is comparable to that of few-layer graphene-reinforced nanocomposites.

**Acknowledgements**

This research has been supported by funding from the European Union Seventh Framework Programme under grant agreement n°604391 Graphene Flagship and the EPSRC (award no. EP/I023879/1). The authors are grateful to Eoghan O'Connell and Ursel Bangert, Bernal Institute, Department of Physics and Energy, University of Limerick, Limerick, Ireland, for help with the TEM analysis.